# Evaluating the Techno-Economic Viability of a Solar PV-Wind Turbine Hybrid System with Battery Storage for an Electric Vehicle Charging Station in Khobar, Saudi Arabia

Ahmed S. AbdElrazek *, Mohamed Soliman *, Muhammad Khalid

*Abstract--* The main aim of this investigation is to replicate and enhance a sustainable hybrid energy structure that combines solar photovoltaic, wind turbines, battery storage. The study employs the Homer simulation model to evaluate the scaling, cost, and control strategy of this hybrid power system. This work primarily focuses on determining the most efficient design for a renewable energy generation system architecture for a significant electric vehicle charging station. The hybrid power system is designed to meet an AC base load of 2424.25 kWh/day with peak consumption of 444 kW. The simulation results indicate that the optimized components and the cost of energy are at an optimal level and the optimal design in terms of renewable energy penetration.

## I. INTRODUCTION

Due to the growing global population, the demand for electricity is rapidly increasing. In actuality, the majority of energy produced today comes from fossil fuels like coal, fuel oil, and natural gas. However, the use of fossil fuels to generate energy results in the emissions of carbon dioxide ($CO_2$), which have a detrimental effect on the environment and cause global warming. Renewable energy sources (RES), such as solar, wind, biomass, and hydroelectric energy, are of tremendous interest for fixing this issue [1]. The fact that electric vehicles (EVs) are becoming more and more popular around the world is evidence of their growing importance. This trend is being fueled by environmental concerns, technological breakthroughs, and a desire for clean, sustainable mobility. The large adoption of EVs depends on the construction of charging stations for them. Electric charge stations may be available in locations away from the grid and need to be supplied with green energy due to international concern for the environment and the minimization of $CO_2$ impact. Microgrid (MG) can be used to supply power to an EV station, and it is an optimal way to feed it with the required electricity [2].

EVs are increasingly popular around the world simultaneous expansion of renewable energy resources (RES) around the world [3]. This circumstance may involve multiple stakeholders. Obtain benefits; these include the EV owner, the end power user, the administrators of the electrical system, and the decision-makers, which are referred to as "stakeholders" combined [4]. Smart charging and Vehicle-to-system technology are used to reduce electricity costs, prevent EV battery deterioration, reduce $CO_2$ emissions, and optimize system performance.

Renewables cannot continually supply and provide the required electricity for urban cities, so the integration of hybrid renewable systems will result in a perfectly reliable and ecologically friendly system. Hybrid renewable systems combine various renewable technologies, such as photovoltaic power, wind power, biomass, and storage systems [5], [6]. The goal of this study is to simulate, optimize and design a system that maximizes the operation of a solar PV and wind hybrid energy system for electrical energy delivery with the lowest energy cost. The system is the main provider of electricity for a specific EV charging station. MGs are now made with the intention of increasing RES penetration. MGs are characterized as small clusters of loads and generation units that have the ability to manage energy locally and operate autonomously [7]. They can also be operated in both isolated and grid-connected modes. Demand and supply imbalance can be perfectly solved with the help of energy storage systems (ESS). Furthermore, the implementation of ESS can enhance the quality and stability of any electrical system.

Maximizing the installed size of battery ESS (BESS) allows for a greater improvement in MG operations and a decrease in the cost of generating from non-renewable sources; however, the high installation costs associated with a wider deployment of BESS pose a challenge. As a result, determining the appropriate BESS size and obtaining the lowest total net present cost (TNPC) for the MG system are necessary in order to minimize the cost of BESS installation [8], [9]. A comparative study between different control approaches will be provided based on the system. A wind-solar hybrid energy charging station was designed and optimized using HOMER software in Western Turkey [10]. Using HOMER software, simulations of an Egyptian MG design with a flywheel ESS have been executed in [11]. After performing an economic analysis of MG with flywheel energy storage, the most cost-effective MG system in "Makkah, Saudi Arabia" was determined Obtaining MG's TNPC in [12].

Ahmed S. AbdErazek is with the Electrical Engineering Department, King Fahd University of Petroleum and Minerals, Dhahran, 31261, Saudi Arabia (e-mails: ahmed.s.razek@gmail.com)

Mohamed Soliman is with the Control & Instrumentation Engineering Department, King Fahd University of Petroleum and Minerals, Dhahran, 31261, Saudi Arabia (e-mails: g202215300@kfupm.edu.sa)

M. Khalid is with the Electrical Engineering Department, King Fahd University of Petroleum and Minerals, Dhahran, 31261, Saudi Arabia and Interdisciplinary Research Center for Sustainable Energy Systems, King Fahd University of Petroleum and Minerals, Dhahran, 31261, Saudi Arabia (e-mail: mkhalid@kfupm.edu.sa)

* Both authors contributed equally

A research study [13] was conducted in "Sohar, Oman" to determine the ideal PV system size, where the reduced TNPC was 6,233 US dollars and the cost of energy (COE) was 0.561 US $/kWh. In [14], the authors suggested using the Homer software to optimize a hybrid RES for rural regions of Tunisia. A variety of energy sources, such as wind/battery, PV/battery, wind/PV/battery, and wind/PV/diesel/battery, are combined in hybrid systems. Consideration is given to both heating and electrical loads. Using HOMER software, an optimal off-grid solution is found for eight situations in the Northeast United Kingdom in [15]. United Kingdom authors in [16] have designed a hybrid energy system with Homer software that combines the production of power and heat. Studies that are relevant to hybrid systems that make use of the Hybrid Optimization Model for Electric Renewable Energy (HOMER).

TABLE I
STUDIES THAT ARE RELEVANT TO HYBRID SYSTEMS THAT MAKE USE OF HOMER SOFTWARE

| Ref. | location | Configuration | LCOE ($/kWh) |
|---|---|---|---|
| [17] | Saudi Arabia | PV-DG-BESS | 0.18 |
| [18] | Nigeria | PV-DG-BESS | 0.41 |
| [19] | Ghana | PV-DG-BESS | 0.53 |
| [20] | Oman | PV-DG-BESS | 0.29 |
| [21] | South Algeria | PV/ DG / DG | 0.940 |
| [22] | South of Algeria | PV/wind/ DG | 0.176 |
| [23] | Southern Norway | Bio/PV/wind/ BESS /capacitor | 0.306 |
| [24] | The northernmost city in Africa | PV/wind/ DG | 0.26 |
| [15] | Northeast United Kingdom | PV/wind/ BESS / biogas | $0.588 |
| [25] | Indonesia | PV/wind/ BESS | 1.06 |
| [26] | Bangladesh | PV/wind/ BESS | 0.47 |
| [27] | Egypt | PV-wind- BESS -grid | 0.114 |
| Present scenario | Saudi Arabia | Grid only | 0.16 |
| Optimal Proposed system | Saudi Arabia | PV-wind- BESS -DG | 0.0955 |

II. HOMER PROGRAM SOFTWARE

HOMER is a simplified version of the acronym hybrid optimization model for electrical renewables. The National Renewable Energy Laboratory (NREL) in the United States came up with and developed it. Homer is a micro power modeling and optimization software that simplifies the process of system evaluation and cost analysis in both grid-connected and stand-alone modes, catering to a wide range of applications. The use of this software is very feasible. The HOMER model was designed to incorporate rate inputs that indicate technology choices, individual constituent expenses, and resource adequacy. HOMER uses these inputs, together with a combination of components, to simulate a variety of system setups. HOMER's simulation results are contained in a register of possible configurations grouped by Net Present Cost (NPC) [28].

A. *Homer Software Dispatch Strategy (Controller)*

The dispatch strategy of the HOMER software specifies how generators and battery banks in a power system should operate. The software offers simulation capabilities for two primary strategies, namely CC (Cost Control) and LF (Load Following). The choice between these strategies depends on various factors, including fuel prices, the generators and battery banks sizes, operation and maintenance (O&M) costs of the generators, renewable resource characteristics, and the amount of renewable energy in the system. The optimal strategy selection is influenced by these factors [29].

- The LF strategy: aims to maintain a balance between the electrical demand and power generation by generators. It ensures that generators only produce the necessary amount of power to meet the immediate demand, allowing the renewable energy sources to handle the charging of storage banks. This method works especially well in systems with a lot of renewable energy sources, especially when those sources periodically produce more power than the demand requires.

- The CC strategy emphasizes maximizing generator utilization by running them at full capacity and using any excess power to recharge the battery bank. This approach proves particularly advantageous in setups with restricted or lacking renewable energy sources, as it efficiently utilizes generator capacity to optimize energy storage for later use.

B. *Homer Software Optimization Criteria*

- Economic

-Total NPC is a comprehensive financial metric that evaluates the overall cost of a system over its lifespan. It considers various factors such as initial capital costs, replacement costs, operational and maintenance (O&M) expenses and fuel expenditures.

-LCOE is a commonly used metric in the energy industry. In HOMER, the LCOE is defined as the average cost per kilowatt-hour (kWh) for the useful electrical power generated by the system. It provides a standardized measure for comparing the cost of electricity generation across different technologies and systems.

- Environmental

HOMER is equipped to calculate the emissions of six pollutants: carbon dioxide ($CO_2$), sulfur dioxide ($SO_2$), nitrogen oxides ($NO_x$), particulate matter (PM), unburned hydrocarbons (UHC), and carbon monoxide (CO). This functionality is essential when planning microgrids with an emphasis on reducing the release of harmful gases. By considering these emissions, HOMER assists in creating eco-friendly and sustainable microgrid configurations.

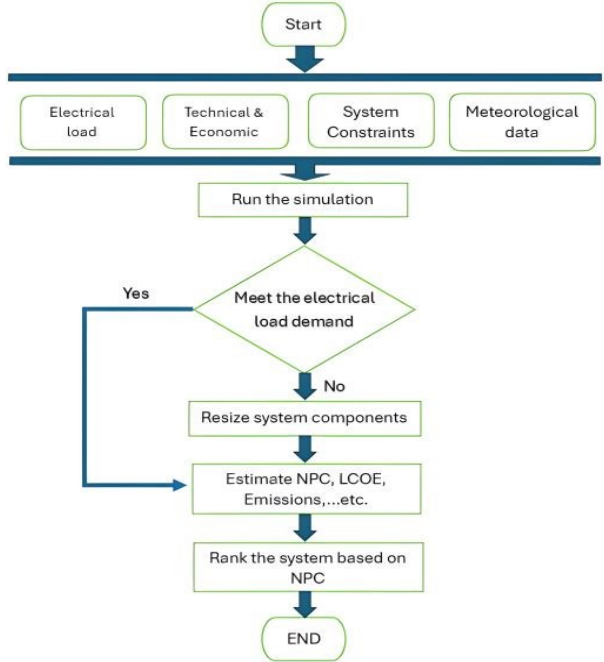

Fig. 1. HOMER methodology flowchart [29].

### III. PROPOSED HYBRID ENERGY SYSTEM

The EV charging station, situated in El Khobar, Saudi Arabia, at coordinates 26.3508° and 50.2123°, serves as a hub for charging EVs. Fig. 2 depicts its precise location, while Fig.3 provides an on-site image of the charging station.

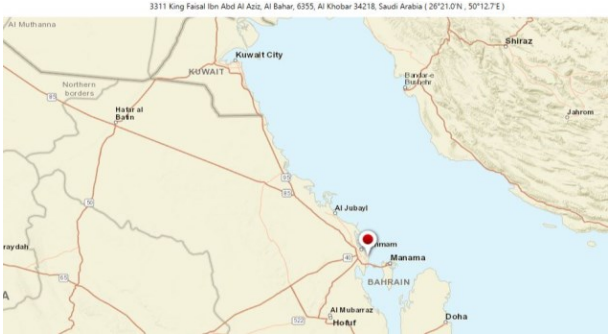

Fig. 2. Geographical position of the EV station

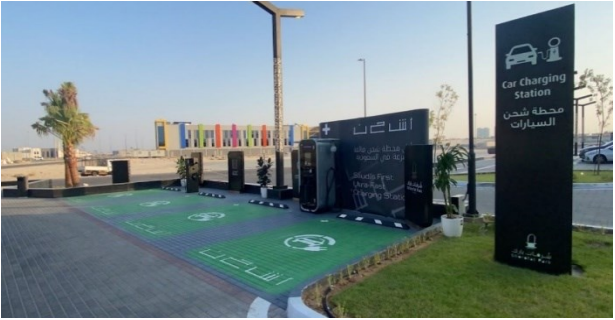

Fig. 3. The EV station of our study

The EV station draws power from the Saudi Arabia grid, where each kilowatt-hour (kWh) costs 0.16 $. Thus, there's a requirement to optimize a hybrid system to supply energy to the EV station at the lowest feasible cost. Additionally, the MG should boast a substantial proportion of renewable sources to curtail emissions. The schematic diagram illustrating the proposed MG for supplying power to EVs is depicted in Fig. 4. It encompasses PV modules, wind turbines, a battery storage system (BSS), an accessible diesel generator, and necessitates a converter.

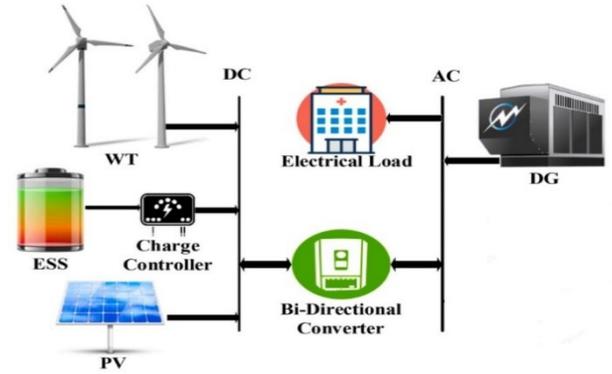

Fig. 4. Schematic diagram for under study MG

A solar photovoltaic energy source should be combined with other energy sources, whether utilized independently or as part of a system. Each hour, Homer Pro calculates the energy flows to and from each system component. The simulation model was created using homer pro software and includes a solar PV, wind turbine, converter, battery storage and diesel generator. Fig. 5 depicts the block diagram for the suggested hybrid energy system in this study.

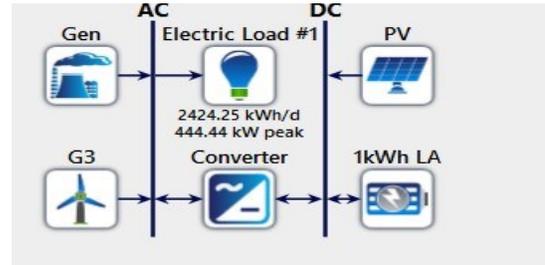

Fig. 5. Hybrid renewable energy system block diagram

### IV. SYSTEM COMPONENT

#### A. Solar Photovoltaic

In order to achieve precise modeling and accurate performance prediction of the PV system, equation (1), is utilized [6], [30]:

$$P_{PV} = Y_{PV} f_{PV} \left(\frac{G_T}{G_{T.STC}}\right)[1 + \alpha_P(T_C - T_{C.STC})] \qquad (1)$$

Where, $f_{PV}$ represents the PV derating factor (%); $Y_{PV}$ denotes the rated capacity of the PV array (kW); $G_T$ is the global solar radiation incident on the surface of the PV array in the current time step (kW/m2); $G_{T.STC} = 1\text{kW/m2}$ is the standard amount of incident radiation at standard test conditions (STC), (25 °C); and $\alpha_T$ is the temperature coefficient of power (% /°C).

Meanwhile, $T_C$ and $T_{C.STC}$ are the PV cell temperature under operating conditions and the PV cell temperature under the STC, respectively. The PV cell temperature was determined by the following equation [31]:

$$T_C = T_a + \left[\frac{(NOCT-20)}{800}\right] * G_T \quad (2)$$

Where $T_a$ represents the surrounding temperature in (◦C), and NOCT is the value of nominal operating cell temperature, which is typically between 45 ◦C and 47 ◦C.

### B. Wind turbine

In HOMER, a wind turbine is modeled as a device that converts wind kinetic energy into AC or DC electricity through a power curve. This curve visually represents the wind turbine's power output in relation to wind speed at the hub height. HOMER's process for estimating wind turbine electricity production involves four main steps. Initially, the software uses wind data to determine the average wind speed for the hour at the anemometer height. Subsequently, it applies either logarithmic or power laws to establish the relationship between the wind speed at the turbine's hub height. The third step utilizes the turbine's power curve to calculate the power output based on standard air density assumptions for a given wind speed. The final step incorporates the air density ratio, which factors in the actual air density compared to standard air density, multiplied by the total power output. To extrapolate wind speed data in HOMER, the following power-law formula is employed [32], [33]:

$$U_{hub} = U_{anem}\left(\frac{z_{hub}}{z_{anem}}\right)^{\alpha} \quad (3)$$

Through the implementation of density correction, power curves generally illustrate wind turbine behavior under standard temperature and pressure conditions (STP). Nevertheless, HOMER introduces modifications to accommodate actual environmental situations. This is accomplished by incorporating the air density ratio into the power value derived from the power curve, taking into consideration the air density at standard temperature and pressure (1.225 kg/m³), in the following manner [34]:

$$P_{WTG} = \left(\frac{\rho}{\rho_0}\right) * P_{WTG.STP} \quad (4)$$

### C. Battery Bank

The following equation indicates the state of charge (SOC) of the BSS depending on the charging and discharging status [35], [36]:

$$SOC(t + \Delta t) = SOC(t)(1 - \sigma_b) + \left(P_{PV}(t) * \eta_{inv} - \frac{P_L(t)}{\eta_{inv}}\right) * \eta_{BC} * \Delta t \quad (5)$$

$$SOC(t + \Delta t) = SOC(t)(1 - \sigma_b) - \left(\frac{P_L(t)}{\eta_{BD}} - P_{PV}(t) * \eta_{inv}\right) * \eta_{BC} * \Delta t \quad (6)$$

Where, $P_L(t)$ is the electrical load demand; $P_{PV}(t)$ indicates the PV output power in kW; $\eta_{inv}$ refers to the efficiency of the inverter (%); $\sigma_b$ is the BSS rate of the self-discharge (%); and $\eta_C$ and $\eta_D$ represent the charging and discharging efficiencies of the BSS (%), respectively. To preserve the operational longevity of ESSs, it is imperative to maintain the SOC within the established maximum $SOC_{max}$ and minimum $SOC_{min}$ thresholds as per the following:

$$SOC_{min} \leq SOC \leq SOC_{max} \quad (7)$$

### D. Converter

Converters are essential in microgrid (MG) systems, facilitating efficient energy transfer among various components. They function as both inverters, converting DC to AC, and rectifiers, converting AC to DC. This capability is especially important in hybrid systems, where seamless energy exchange between the DC and AC components is vital for ensuring system stability and reliability. The capacity and limitations of the inverter, denoted as $P_{inv}$, are specified in the following equations [37]:

$$P_{inv} = \frac{E_L(max)}{\eta_{dc/ac}} \quad (8)$$

$$\eta_{dc/ac}[P_{BSS.ch/dis}(t) + P_{PV}(t)] \leq P_{inv} \quad (9)$$

$$\eta_{dc/ac}[P_G(t)] \leq P_{inv} \quad (10)$$

The converter capacity is noted by $P_{inv}$ in kW and the maximum electrical load demand $E_L(max)$ in Wh. It also considers the converter efficiencies for DC to AC (ηdc/ac) and AC to DC (ηac/dc) current conversions. Additionally, it accounts for the charge/discharge of the BESS denoted as $P_{BSS.ch/dis}$ as well as the PV power (PPV) and import/export grid powers (PG).

## V. MODELL INPUT INFORMATION

### A. Primary load information

The electrical demand in this context is referred to as the load. The energy demand was estimated based on one of the EV Charging Stations. In the system architecture, the first step involved entering the hourly load data into the HOMER software. To handle the large-scale load information from Khobar– Saudi Arabia electricity consumption at the EV Charging Station, the daily and monthly load profile can be observed in Figs 6, 7. The annual average of the system is 2424.2 kWh/day, with a peak load of 390.41 kWh. The load factor is calculated to be 0.26.

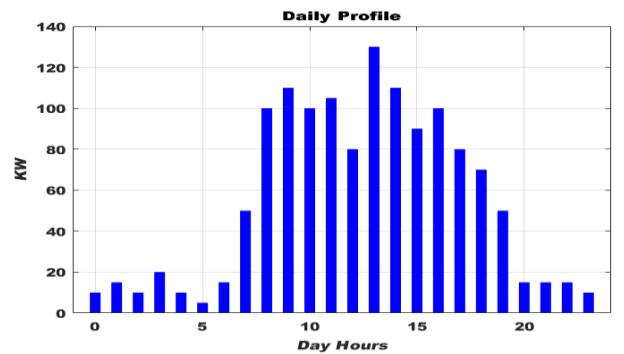

Fig. 6. The average daily load profile of the selected case study.

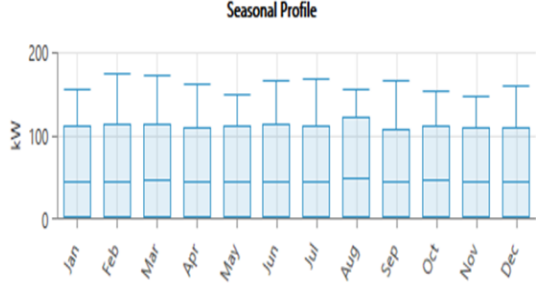

Fig. 7. The monthly load profile of the selected case study.

### B. Solar radiation information

The solar irradiance data for a specific area of interest was obtained from the NASA Surface meteorology and Solar Energy database. Analysis of the data revealed that the scaled annual average solar irradiance is 5.6 kWh/m2/d. Additionally, in the region of Khobar, Saudi Arabia, an EV charging station was established and operates monthly. This information is presented in Fig. 8. Moreover, the daily temperature is presented in Fig. 9.

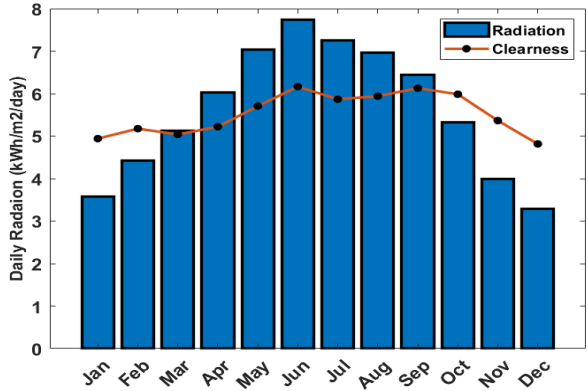

Fig. 8. Annual average global irradiance of the selected study

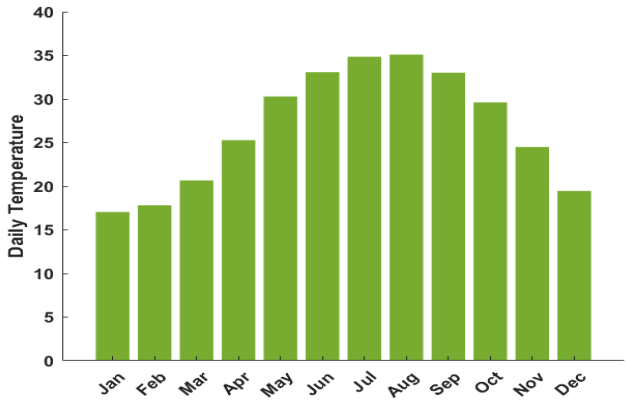

Fig. 9 Daily Temperature in The Selected place

### C. Wind speed information

The wind energy data for a particular region of interest has been provided by NASA surface meteorology. The data shows a scaled average annual wind speed of 5.61 m/s. Fig. 9 illustrates the monthly wind speed for an EV Charging Station in Khobar–Saudi Arabia.

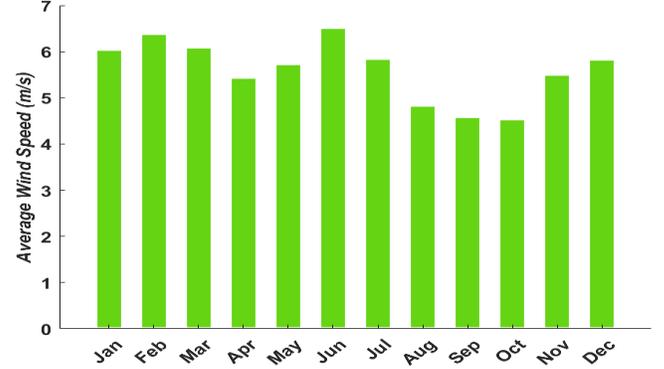

Fig. 10. Annual average wind speed of the selected case study

## VI. RESULTS AND DISCUSSION

EV station currently draws power from the main grid, which supplies energy at a Levelized Cost of Electricity (LCOE) of $0.16 per kilowatt-hour (kWh), with no renewable energy contribution. The economic analysis of this scenario includes annual operating costs, along with their equivalents in terms of total net present cost, as outlined in Table II.

TABLE II
TECHONOMIC AND ENVIRONMENTAL ANALYSIS OF CURRENT SCENARIO

| Term | Value | Units |
|---|---|---|
| Operating cost | 141,576 | $/year |
| Capital cost | 0 | $ |
| NPC | 1.8 | M$ |
| LCOE | 0.16 | $/kWh |
| Renewable Factor | 0 | % |
| Carbon dioxide | 559,226 | Kg/year |
| Sulfur Dioxide | 2,424 | Kg/year |
| Nitrogen Oxides | 1,186 | Kg/year |

In this study, the HOMER simulation model was utilized to investigate a hybrid energy system that integrates solar PV, wind turbine, battery, diesel generator, and converter. The analysis yielded results pertaining to the system architecture, cost summary, and electrical aspects, which are elaborated upon below. The details of the components used in the system are described as follows in Table III.

TABLE III
COMPONENTS DETAILS OF SUGGESTED SYSTEM ARCHITECTURE

| Component | Parameter | Value | Unit | Ref |
|---|---|---|---|---|
| PV (Trina Solar) | Rated Capacity | 1 | kW | [28] |
| | Temperature coefficient | -0.34 | %/°C | |
| | Operating temperature | 43 | °C | |
| | Panel lifetime | 25 | Year | |
| | Capital cost | 222.7 | $/kW | |
| | O&M cost | 4.45 | $/year | |
| Wind turbine | Rated Capacity | 3 | kW | [17] |
| | Tower height | 12 | m | |
| | Capital Cost | 1086 | $/unit | |
| | O&M cost | 44 | $/year | |
| Battery (Power Safe SBS F150) | Rated Capacity | 2 | kWh | [38] |
| | Nominal Voltage | 12 | V | |
| | Round trip efficiency | 97 | % | |

| | Capital Cost | 200 | $/unit | |
|---|---|---|---|---|
| | Replacement cost | 200 | $/unit | |
| | O&M cost | 2 | $/year | |
| | Life time | 5 | Year | |
| | Max depth of discharge | 100 | % | |
| | Min depth of discharge | 20 | % | |
| DG (Autosize Genset) | Capital Cost | 500 | $/kW | [28] |
| | Replacement cost | 500 | $/kW | |
| | O&M cost | 0.03 | $/op.h | |
| | Life time | 15000 | h | |
| | Fuel cost | 0.168 | $/L | |
| Converter (Schneider) | Power Capacity | 1 | kW | |
| | Capital Cost | 280 | $/kW | |
| | Replacement cost | 280 | $/kW | |
| | O&M cost | 10 | $/year | |
| | Efficiency | 97 | % | |
| | Lifetime | 25 | Year | |
| Project | Lifetime | 25 | Year | [39] |
| | Nominal discount rate | 8.12 | % | |
| | Real discount rate | 6 | % | |
| | Expected inflation rate | 2 | % | |

The simulation results indicate two potential scenarios for the optimal setup of EV stations: one focused on achieving the lowest cost of energy, while the other aims for 100% penetration of renewable energy.

*A. First scenario*

The first configuration consists of PV generators and WT turbines, a BSS, and a diesel generator to obtain the lowest cost of electricity with LCOE equal to 0.0955$/kWh. The number of units of each source and the economic analysis of this configuration are detailed in Table IV. Table V shows the environmental analysis of the first scenario according produced quantities of each pollutants.

TABLE IV
FIRST SCENARIO DESIGN AND ECONOMICS

| Parameter | Value | Units |
|---|---|---|
| $N_{PV}$ (Number of PV units) | 714 | Unit |
| $N_{WT}$ (Number of WT units) | 67 | Unit |
| $N_{BT}$ (Number of BT units) | 1059 | Unit |
| DG | 490 | kW |
| Converter | 331 | kW |
| Operating cost | 31,667 | $/year |
| Capital cost | 675,189 | $ |
| NPC | 1.08 | M$ |
| LCOE | 0.0955 | $/kWh |
| Renewable Factor | 95.8 | % |

TABLE V
FIRST SCENARIO ENVIRNOMENTAL ANALYSIS

| Term | Value | Units |
|---|---|---|
| Carbon dioxide | 29,530 | Kg/year |
| Carbon Monoxide | 186 | Kg/year |
| Unburned Hydrocarbons | 8.12 | Kg/year |
| Particular Matter | 1.13 | Kg/year |
| Sulfur Dioxide | 72.3 | Kg/year |
| Nitrogen Oxides | 175 | Kg/year |

Table VI presents the annualized expenses of the system in the first scenario, comprising initial investment outlays, replacement expenditures, operational and maintenance (O&M) outlays, and the residual values for every element. Additionally, Fig. 11 illustrates a comparison of the contributions of each technology to the overall costs.

TABLE VI
COST DETAILS OF THE OPTIMAL SYSTEM (FIRST SCENARIO)

| Component | Capital ($) | Replacement ($) | O&M ($) | Fuel ($) | Salvage ($) | Total ($) |
|---|---|---|---|---|---|---|
| DG | 245,000 | 0 | 51,865 | 24,228 | 30,826 | 290,266 |
| BT | 105,900 | 215,477 | 13,538 | 0 | 0 | 334,915 |
| WT | 72,762 | 22,687 | 37,685 | 0 | 12,715 | 120,419 |
| PV | 158,915 | 0 | 40,593 | 0 | 0 | 199,508 |
| Converter | 92,612 | 0 | 42,282 | 0 | 0 | 134,894 |
| System | 675,189 | 238,165 | 185,962 | 24,228 | 43,541 | 1,080,001 |

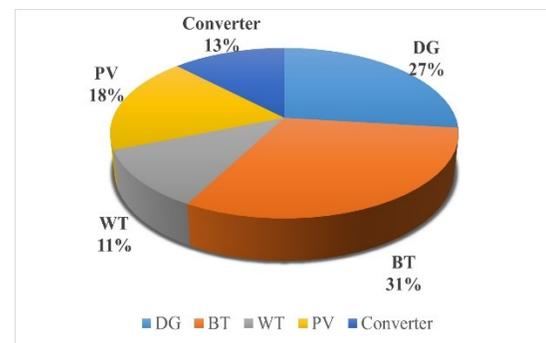

Fig. 11. Cost sharing between used technologies in case 1

Fig. 12 displays the total system generation for each month, along with the contribution of each energy source in every month.

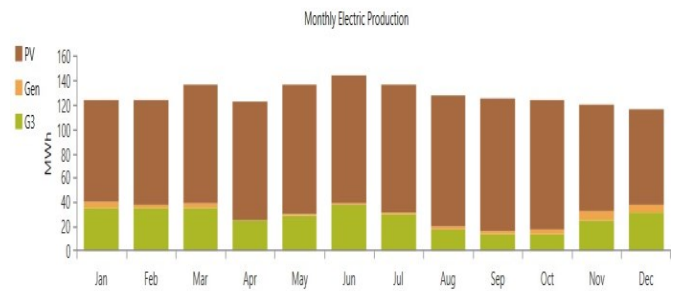

Fig. 12. The monthly average energy production from the proposed system

*B. Second scenario*

The second system uses solar panels, a battery storage system, and a diesel generator to produce electricity at a cost of $0.1073 per kilowatt-hour. The specific number of each type of equipment and the financial analysis of this system are presented in Table VII. Table VIII shows the environmental impact of the second system, including the amount of each pollutant produced. The second configuration and system solar panels, a battery storage system, and a diesel generator and has a LCOE equal 0.1073 $/kWh. Which is making saving around 32.94% than the current scenario, but it still higher than the first scenario by around 12.235 %.

TABLE VII
SECOND SCENARIO DESIGN AND ECONOMICS

| Parameter | Value | Units |
|---|---|---|
| $N_{PV}$ (Number of PV units) | 918 | Unit |
| $N_{BT}$ (Number of BT units) | 1414 | Unit |
| DG | 490 | kW |
| Converter | 372 | kW |
| Operating cost | 40,620 | $/year |
| Capital cost | 695,002 | $ |
| NPC | 1.21 | M$ |
| LCOE | 0.1073 | $/kWh |
| Renewable Factor | 94.5 | % |

TABLE VIII
SECOND SCENARIO ENVIRNOMENTAL ANALYSIS

| Term | Value | Units |
|---|---|---|
| Carbon dioxide | 37,923 | Kg/year |
| Carbon Monoxide | 239 | Kg/year |
| Unburned Hydrocarbons | 10.4 | Kg/year |
| Particular Matter | 1.45 | Kg/year |
| Sulfur Dioxide | 92.9 | Kg/year |
| Nitrogen Oxides | 225 | Kg/year |

Table IX presents the annualized expenses of the system in the second scenario, comprising initial investment outlays, replacement expenditures, operational and maintenance (O&M) outlays, and the residual values for every element. Additionally, Fig. 13 illustrates a comparison of the contributions of each technology to the overall costs for the second scenario.

TABLE IX
COST DETAILS OF THE OPTIMAL SYSTEM (SECOND SCENARIO

| component | Capital ($) | Replacement ($) | O&M ($) | Fuel ($) | Salvage ($) | Total ($) |
|---|---|---|---|---|---|---|
| DG | 245,000 | 0 | 65,582 | 31,113 | 23,880 | 317,815 |
| BT | 141,400 | 343,813 | 18,076 | 0 | 15,217 | 448,071 |
| PV | 204,462 | 0 | 52,227 | 0 | 0 | 256,689 |
| Converter | 104,140 | 0 | 47,545 | 0 | 0 | 151,685 |
| System | 695,002 | 343,813 | 183,430 | 31,113 | 39,098 | 1,214,261 |

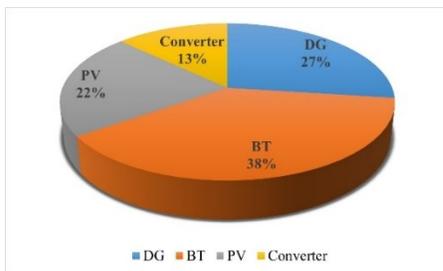

Fig.13. Cost sharing between used technologies in scenario 2

### C. Third scenario

The third configuration consists of PV, wind turbine, a BSS, to obtain with LCOE equal to 0.124 $/kWh, the number of units of each source, and the economic analysis of this configuration is detailed in Table X. This case is dependent on renewable energy totally 100 %, so there are no emissions of pollutants to the air while producing this electricity. The number of units of each source and the economic analysis of this configuration are detailed in Table XI.

The third and fourth scenarios are totally 100 % renewable fraction. The third configuration contain solar panels, a battery storage system, and a diesel generator has a LCOE equal 0.124 $/kWh which make a saving a round 22.5 % than the current scenario. But it higher than the first scenario and second scenario by about 29.84 % and 15.56 % respectively however is highlighted than the first and scenarios as it is totally 100 % renewable fraction unlike the first scenario.

TABLE X
THIRD SCENARIO DESIGN AND ECONOMICS

| Parameter | Value | Units |
|---|---|---|
| $N_{PV}$ | 2188 | unit |
| $N_{BT}$ | 1618 | Unit |
| $N_{WT}$ | 51 | kW |
| Converter | 464 | kW |
| Operating cost | 44,585 | $/year |
| Capital cost | 843,369 | $ |
| NPC | 1.4 | M$ |
| LCOE | 0.124 | $/kWh |
| Renewable Factor | 100 | % |

TABLE XI
COST DETAILS OF THE THIRD SCENARIO

| component | Capital ($) | Replacement ($) | O&M ($) | Salvage ($) | Total ($) |
|---|---|---|---|---|---|
| BT | 161,800 | 329,218 | 20,683 | 0 | 511,702 |
| WT | 55,386 | 17,269 | 28,686 | 9,679 | 91,663 |
| PV | 487,323 | 0 | 124,481 | 0 | 611,804 |
| Converter | 129.860 | 0 | 59,287 | 0 | 189,147 |
| System | 834,369 | 346,488 | 223,137 | 9,679 | 1,404,316 |

Fig. 14 illustrates a comparison of the contributions of each technology to the overall costs for the third scenario

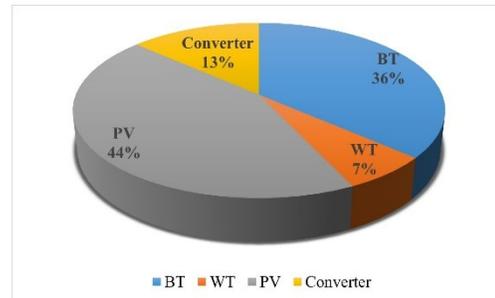

Fig.14. Cost sharing between used technologies third scenario

### D. Fourth scenario

The last configuration consists of PV and a BSS, to obtain with LCOE equal to 0.142 $/kWh, the number of units of each source, and the economic analysis of this configuration is detailed in Table XII. This case is dependent on renewable energy totally 100 %, so there are no emissions of pollutants into the air while producing this electricity. The number of units of each source and the economic analysis of this configuration are detailed in Table XIII. While the fourth configuration and solution which contains only PV modules

and batteries can achieve 100% penetration of renewable energy, the cost of electricity will be $0.142 per kWh, resulting in savings of approximately 11.25 % compared to the current scenario. However, this cost is higher than that of scenario 1 by 48.69 %. Scenario 4 incurs no emissions, thereby reducing carbon dioxide and nitrogen oxides emissions by 100%. LCOE in Scenario four is also higher than it in scenario 2 and 3 by around 32.34 %, 14.15 %, respectively.

TABLE XII
THIRD SCENARIO DESIGN AND ECONOMICS

| Parameter | Value | Units |
|---|---|---|
| $N_{PV}$ (Number of PV units) | 2867 | Unit |
| $N_{BT}$ (Number of BT units) | 1936 | Unit |
| Converter | 456 | kW |
| Operating cost | 50,071 | $/year |
| Capital cost | 959,799 | $ |
| NPC | 1,6 | M$ |
| LCOE | 0.142 | $/kWh |
| Renewable Factor | 100 | % |

TABLE XIII
COST DETAILS OF THE THIRD SCENARIO

| component | Capital ($) | Replacement ($) | O&M ($) | Salvage ($) | Total ($) |
|---|---|---|---|---|---|
| BT | 193,600 | 393,966 | 24748 | 0 | 612,271 |
| PV | 638,456 | 0 | 163,086 | 0 | 801,542 |
| Converter | 127,742 | 0 | 58,320 | 0 | 186,062 |
| System | 959798 | 393,922 | 246,155 | 0 | 1,599,876 |

Fig. 15 illustrates a comparison of the contributions of each technology to the overall costs for the fourth scenario.

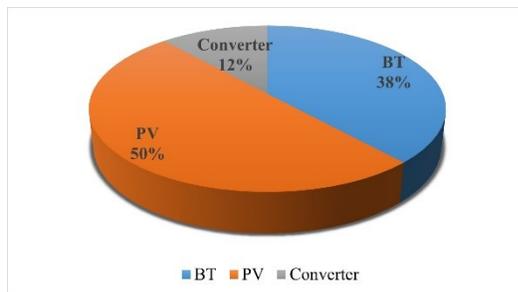

Fig.15. Cost sharing between used technologies fourth scenario

## VII. CONCLUSION

This study demonstrates the simulation and optimization of a renewable hybrid energy system combining solar PV and wind turbines for an electric vehicle (EV) charging station in Khobar, Saudi Arabia. The analysis, conducted using HOMER software, identifies the most economically viable configuration: 714 solar panels, 67 wind turbines, a 331-kW converter, 1059 batteries, and a 490-kW diesel generator. This setup yields a total net present cost of $1.08 million, with a cost of energy (COE) of $0.0955, and significantly reduces operating costs and emissions, achieving a renewable fraction of 95.8%. While 100% renewable configurations (scenario 3 and scenario 4) were also explored, they incurred higher COEs. Scenario 4, which relied solely on solar PV and batteries, achieved 100% renewable penetration but had the highest COE. Overall, the study demonstrated the economic viability and environmental benefits of hybrid renewable energy systems for EV charging stations in Saudi Arabia. These systems can effectively meet load demands and contribute to the country's renewable energy goals. Future research can further explore the optimization of such systems and their integration into broader energy grids.

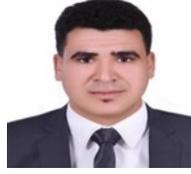
**Ahmed S. AbdElrazek** received his bachelor's and master's degrees in electrical engineering in 2018 and 2023, respectively. He is currently a PhD student in the electrical engineering department at King Fahd University of Petroleum and Minerals (KFUPM), Dhahran, Saudi Arabia. He has many scientific research articles published in prestigious international journals and conferences. His research interests include hybrid systems, microgrids, power systems, stability, power electronics, photovoltaics, wind energy, and meta-heuristic optimization algorithms.

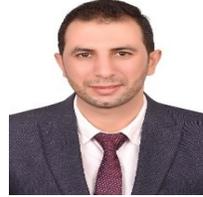
**Mohamed Soliman** obtained his bachelor's and master's degrees in automatic control engineering from Menoufia University, Shibin Al Kawm, Egypt, in 2015 and 2021, respectively. Currently, he is working towards his Ph.D. in intelligent control systems at King Fahd University for Petroleum and Minerals (KFUPM) and holds a position as a Teaching Assistant in the Department of Control and Instrumentation Engineering (CIE) at KFUPM. His research focus encompasses intelligent control systems, robust control, adaptive control, optimization algorithms, and nonlinear control systems.

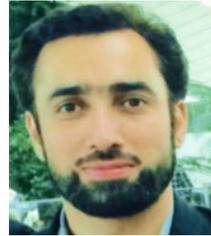
**Muhammad Khalid** (Senior Member, IEEE) received the Ph.D. degree in electrical engineering from the School of Electrical Engineering Telecommunications (EET), University of New South Wales (UNSW), Sydney, Australia, in 2011. He was a Postdoctoral Research Fellow for three years and then he continued as a Sr. Research Associate with the Australian Energy Research Institute, School of EET, UNSW, for another two years. He is currently an Associate Professor with the Electrical Engineering Department, King Fahd University of Petroleum and Minerals (KFUPM), Dhahran, Saudi Arabia. He has authored/co-authored several journal and conference papers in the field of control and optimization for renewable power systems. His current research interests include the optimization and control of battery energy storage systems for large-scale grid connected renewable power plants (particularly wind and solar), distributed power generation and dispatch, hybrid energy storage, hydrogen systems, EVs, and smart grids. He was a recipient of a highly competitive postdoctoral writing fellowship from UNSW, in 2010. He was a recipient of many academic awards and research fellowships. He has been a reviewer for numerous international journals and conferences.